\documentstyle[aps,12pt]{revtex}

\begin{document}
\draft
\author{S. N. Dolya$^{1}$ and O. B. Zaslavskii$^{2}$}
\address{$^{1}$B. Verkin Institute for Low Temperature Physics and Engineering, 47\\
Lenin Prospekt, Kharkov 61164, Ukraine\\
E-mail: dolya@ilt.kharkov.ua\\
$^{2}$Department of Physics, Kharkov V. N. Karazin's National University,\\
Svoboda Sq.4, Kharkov 61077, Ukraine\\
E-mail: aptm@kharkov.ua}
\title{Quantum anharmonic oscillator and quasi-exactly solvable Bose systems }
\maketitle

\begin{abstract}
We extend the notion of quasi-exactly solvable (QES) models from potential
ones and differential equations to Bose systems. We obtain conditions under
which algebraization of the part of the spectrum occurs. In some particular
cases simple exact expressions for several energy levels of an anharmonic
Bose oscillator are obtained explicitly. The corresponding results do not
exploit perturbation theory and include strong coupling regime. A\ generic
Hamiltonian under discussion cannot, in contrast to QES potential models, be
expressed as a polynomial in generators of $sl_{2}$ algebra. The suggested
approach is extendable to many-particle Bose systems with interaction.
\end{abstract}



\section{introduction}

An anharmonic oscillator represents one of the ''eternal'' problems and
models of theoretical physics. It serves as a basis for checking different
approximate methods in quantum mechanics, the simplified counterpart of
field-theoretical models, etc. Apart from this, it is of interest on its own
since the real world certainly deviates from idealized picture of harmonic
oscillators due to interaction between them and self-interaction. In so
doing, the notion ''anharmonic oscillator'' is mainly applied at least to
two different entities. First, it refers to some potential power models in
which the potential contains terms with the higher degrees with respect to
coordinate. (The literature on this subject is so vast, that it is even
rather difficult to indicate some concrete references - let us mention here
only the reviews \cite{wu}, \cite{turb89}, the book \cite{klein} and
references therein). Second, it is related to quantum Bose models with
interaction or self-interaction. In both cases the Schr\"{o}dinger equation
cannot be solved exactly. However, for the first case it was realized that,
in spite of impossibility to find the whole energy spectrum exactly, in some
particular cases (for instance, sextic oscillator with special relationship
between coefficients \cite{ush}) one can find the {\it part} of the spectrum
(more precisely, algebraization of the part of the spectrum occurs). Such a
system is the example of so-called quasi-exactly solvable (QES) \cite{hist}
ones which includes a rather vast class of potentials and have direct
physical meaning, first of all related to properties of magnetic systems 
\cite{uz}.

The aim of the present paper is to extend the notion of QES systems to Bose
ones and apply QES approach to anharmonic Bose oscillators. Strange as it
may seem, the approach to Bose oscillators in the spirit of QES models was,
to the best of our knowledge, absent in literature before in spite of the
developed apparatus relating realization of Lie algebras in Fock space and
properties of differential equations \cite{turb97}. Meanwhile, the QES
approach to Bose systems deserves treatment on its own due to an obviously
wide area of physical applications.

In the coordinate-momentum representation 
\begin{equation}
a\rightarrow \frac{d}{dx}\text{, }a^{+}\rightarrow x\text{,}  \label{a}
\end{equation}
Bose Hamiltonian, polynomial with respect to $a$, $a^{+}$ becomes a
differential operator and any possible invariant subspace is spanned on the
polynomial basis. For the systems of such a kind there exists the Turbiner's
theorem that states that the most general QES can be expressed in terms of
generators of $sl_{2}$ algebra 
\begin{equation}
J^{+}=a^{+2}a-Na^{+}\text{, }J^{-}=a\text{, }J^{0}=a^{+}a-\frac{N}{2}\text{.}
\label{j}
\end{equation}
(the relations (\ref{j}) are known in the magnetism theory as the
Dayson-Maleev representation \cite{dm}). Nevertheless, the problem of
finding QES Bose Hamiltonians cannot be exhausted by a simple reference to
this theorem since, as we will below, for the typical case under discussion,
the conditions of validity of this theorem are not fulfilled, so
Hamiltonians (except some special cases) cannot be expressed in terms of $%
sl_{2}$ generators at all. This is the point in which Bose QES systems
qualitatively differ from potential QES models whose Hamiltonians are built
with the help of $sl_{2}$ generators, realized as differential operators 
\cite{hist}.

Apart from this, even in the cases when Turbiner's theorem does apply to
Bose Hamiltonians, it is much more convenient to formulate the conditions of
QES - solvability in terms of coefficients of an original Bose Hamiltonian
directly without resorting to operators $J^{i}$ at the intermediate stage.

\section{Basic formulas}

Consider Hamiltonian 
\begin{equation}
H=H_{0}+V\text{, }H_{0}=\sum_{p=1}^{p_{0}}\varepsilon _{p}(a^{+}a)^{p}\text{%
, }V=\sum_{s=0}^{s_{0}}A_{s}[(a^{+}a)^{s}a^{2}+(a^{+})^{2}(a^{+}a)^{s}]\text{%
.}  \label{h}
\end{equation}
Throughout the paper we assume that all coefficients of Hamiltonian are
real. For Hamiltonian (\ref{h}) to have a well-defined ground state, one
should take $p_{0}>s_{0}+2$ independently of the relations between
coefficients or $p_{0}=s_{0}+2$ provided $\varepsilon _{p_{0}}\geq
2A_{s_{0}} $. In the $x$-representation (\ref{a})

we obtain 
\begin{equation}
H_{x}=\sum_{p=1}^{p_{0}}\varepsilon _{p}(x\frac{d}{dx})^{p}+%
\sum_{s=0}^{s_{0}}A_{s}[(x\frac{d}{dx})^{s}\frac{d^{2}}{dx^{2}}+x^{2}(x\frac{%
d}{dx})^{s}]\text{.}  \label{hs}
\end{equation}
We are interested in the solutions of Schr\"{o}dinger equation of the type $%
\mid \psi \rangle =\sum_{n=0}^{N}b_{n}\mid n\rangle $, where $\mid n\rangle $
is the state with $n$ particles: $a^{+}a\mid n\rangle =n\mid n\rangle $. For
Hamiltonian (\ref{hs}) subspaces with even and odd are not mixed. Therefore,
it makes sense to consider them separately. In $x$ representation (\ref{a})
the wave function of even states $\Phi =\sum_{l=0}a_{l}\Phi _{l}$, $\Phi
_{l}\equiv x^{2l}$.

It follows from (\ref{hs}) that

\begin{eqnarray}
H_{x}\Phi _{l} &=&\alpha _{l}\Phi _{l+1}+\beta _{l}\Phi _{l-1}+\gamma
_{l}\Phi _{l}\text{, }l=0\text{,}1\text{,..}L\text{,}  \label{hf} \\
\alpha _{l} &=&\sum_{s=0}^{s_{0}}A_{s}(2l)^{s}\text{,}  \nonumber \\
\beta _{l} &=&\sum_{s=0}^{s_{0}}A_{s}2l(2l-1)(2l-2)^{s}\text{,}  \nonumber \\
\gamma _{l} &=&\sum_{p=1}^{p_{0}}\varepsilon _{p}(2l)^{p}\text{.}  \nonumber
\end{eqnarray}
We are interested in the possibility of the existence of the invariant basic 
$F_{2L}=\{1,x^{2},x^{4}...x^{2L}\}$. The condition of cut off at $l=L$
reads: $\alpha _{L}=0$.

For odd states the invariant basic $F_{2M+1}=\{x$, $x^{3}$,... $x^{2M+1}\}$, 
$\tilde{\Phi}_{m}=x^{2m+1}$ and 
\begin{eqnarray}
H_{x}\tilde{\Phi}_{m} &=&\tilde{\alpha}_{m}\tilde{\Phi}_{m+1}+\tilde{\beta}%
_{m}\tilde{\Phi}_{m-1}+\tilde{\gamma}_{m}\tilde{\Phi}_{m}\text{, }m=0,1,...M%
\text{,}  \label{odd} \\
\tilde{\alpha}_{m} &=&\sum_{s=0}^{s=s_{0}}A_{s}(2m+1)^{s}\text{,}  \nonumber
\\
\tilde{\beta}_{m} &=&\sum_{s=0}^{s=s_{0}}A_{s}(2m+1)2m(2m-1)^{s}\text{,} 
\nonumber \\
\tilde{\gamma}_{m} &=&\sum_{p=1}^{p=p_{0}}\varepsilon _{p}(2m+1)^{p}\text{.}
\nonumber
\end{eqnarray}

The subspace with $m\leq M$ is invariant with respect to the action of $H$
provided $\tilde{\alpha}_{M}=0$.

The procedure described above is, in fact, nothing else than the Bose
version of quasi-exactly solvable (QES) models, applied to an anharmonic
oscillator. Now we would like to point out why for the case under
consideration Turbiner's theorem, in general, does not hold, so our formulas
cannot considered as particular cases of its realization. The point is that
Turbiner's theorem implies that the space of {\it all} polynomials of a
given degree is invariant with respect to $J^{i}$: $F_{N}=\{1$, $x$, $x^{2}$%
, ...$x^{N}\}$. Meanwhile, in our case, only subset $F_{2N}$ (for even
states) or $F_{2M+1}$ (for odd ones) is invariant, whereas the set $F_{N}$
is not. Only in particular cases, when both conditions $\alpha _{L}=0$ (for
even states) and $\tilde{\alpha}_{M}=0$ (for odd states) are satisfied
simultaneously, Hamiltonian does become an algebraic combinations of $J^{i}$.

In contrast to \cite{cont}, where differential equations was the object of
research, in our paper the coordinate-momentum representation (\ref{a}), in
which the operator $a$ becomes differential, is used as an useful device at
an intermediate stage only. In principle, one could rely directly on the
known formulas of the action of operators $a$, $a^{+}$ on a states with a
definite number of particles without resorting to the representation (\ref{a}%
).

\section{examples}

Consider Hamiltonain whose off-diagonal part reads

\begin{equation}
V=A_{0}(a^{2}+a^{+2})+A_{1}[(a^{+}a)a^{2}+a^{+2}(a^{+}a)]+A_{2}[(a^{+}a)^{2}a^{2}+a^{+2}(a^{+}a)^{2}]%
\text{, }s_{0}=2\text{.}  \label{h1}
\end{equation}
Now 
\begin{eqnarray}
\alpha _{l} &=&A_{0}+2lA_{1}+(2l)^{2}A_{2}\text{, }  \label{coef} \\
\beta _{l} &=&2l(2l-1)[A_{0}+(2l-2)A_{1}+(2l-2)^{2}A_{2}]\text{.}  \nonumber
\end{eqnarray}
\begin{eqnarray}
\tilde{\alpha}_{m} &=&A_{0}+(2m+1)A_{1}+(2m+1)^{2}A_{2}\text{,}  \label{o2}
\\
\tilde{\beta}_{m} &=&(2m+1)2m[A_{0}+(2m-1)A_{1}+(2m-1)^{2}A_{2}]\text{.} 
\nonumber
\end{eqnarray}
First, consider even states. In the simplest nontrivial particular case the
invariant subspace is two-dimensional, $L=1$. Then 
\begin{equation}
\alpha _{1}=A_{0}+2A_{1}+4A_{2}=0\text{,}  \label{a2}
\end{equation}
$\Phi =a_{0}\Phi _{0}+a_{1}\Phi _{1}$ and it follows from the
Schr\"{o}dinger equation $H\Phi =E\Phi $ that $-Ea_{0}+\beta _{1}a_{1}=0$, $%
\alpha _{0}a_{0}+(\gamma _{1}-E)a_{1}=0$. Taking also into account (\ref{a2}%
), we obtain: 
\begin{equation}
E=\frac{\gamma _{1}}{2}\pm \sqrt{\frac{\gamma _{1}^{2}}{4}%
+8(A_{1}+2A_{2})^{2}}  \label{e2}
\end{equation}

In a similar way, one gets for the three-dimensional subspace ($L=2)$: 
\begin{equation}
E^{3}-(\gamma _{1}+\gamma _{2})E^{2}+[\gamma _{1}\gamma
_{2}-16(5A_{1}^{2}+52A_{1}A_{2}+140A_{2}^{2})]E+32\gamma
_{2}(A_{1}+4A_{2})^{2}=0\text{.}  \label{e3}
\end{equation}
If $A_{1}=A_{2}=0$, the low-lying energy levels of an harmonic oscillator
are reproduced from (\ref{e2}), (\ref{e3}). The equation (\ref{e3}) can be
solved exactly in the particular case $A_{1}=-4A_{2}$: $E=0$, $\frac{\gamma
_{1}+\gamma _{2}}{2}\pm \sqrt{\frac{(\gamma _{1}-\gamma _{2})^{2}}{4}%
+192A_{2}^{2}}$. For odd states in the simplest nontrivial case $M=1$ we
have 
\begin{eqnarray}
\tilde{\alpha}_{1} &\equiv &A_{0}+3A_{1}+9A_{2}=0\text{,}  \label{od2} \\
\tilde{\alpha}_{0} &=&A_{0}+A_{1}+A_{2}\text{, }\tilde{\beta}_{1}=6\tilde{%
\alpha}_{0}\text{,}  \nonumber \\
E &=&\frac{\tilde{\gamma}_{0}+\tilde{\gamma}_{1}}{2}\pm \sqrt{\frac{(\tilde{%
\gamma}_{0}+\tilde{\gamma}_{1})^{2}}{4}+\tilde{\alpha}_{0}\tilde{\beta}_{1}}=%
\frac{\tilde{\gamma}_{0}+\tilde{\gamma}_{1}}{2}\pm \sqrt{\frac{(\tilde{\gamma%
}_{0}+\tilde{\gamma}_{1})^{2}}{4}+24(A_{1}+4A_{2})^{2}}\text{.}  \nonumber
\end{eqnarray}

The conditions $\alpha _{L}=0$ and $\tilde{\alpha}_{M}=0$ are different, so
in general the invariant subspace exists only for even or only for odd
states. However, it may happen that both conditions are fulfilled. Thus, for 
$L=1=M$ the compatibility of (\ref{a2}) and (\ref{od2}) demands $%
A_{1}=-5A_{2}$, $A_{0}=6A_{2}$. Then we have the simple explicit solutions
for 4 levels of Hamiltonian (\ref{h}): $E=\frac{\tilde{\gamma}_{0}+\tilde{%
\gamma}_{1}}{2}\pm \sqrt{\frac{(\tilde{\gamma}_{0}+\tilde{\gamma}_{1})^{2}}{4%
}+24A_{2}^{2}}$, $\frac{\gamma _{1}}{2}\pm \sqrt{\frac{\gamma _{1}^{2}}{4}%
+72A_{2}^{2}\text{.}}$

\section{explicit solution for two levels}

If the coefficient $\alpha _{L}=0$, the dimension of the invariant space is $%
L+1$. Meanwhile, it may happen that, in addition, $\beta _{L-1}=0$. Then the
two-dimensional subspace spanned on $\Phi _{L}$ and $\Phi _{L-1}$ is singled
out from the $L+1$ subspace that gives explicit simple exact solutions for
two levels, whatever large $L$ would be. For Hamiltonian (\ref{h1}) it
follows from (\ref{coef}) that in this case 
\begin{eqnarray}
A_{0} &=&2L(2L-3)A_{2}\text{, }A_{1}=A_{2}(3-4L)\text{,}  \label{cut} \\
\alpha _{L-1} &=&-2A_{2}\text{, }\beta _{L}=-4L(2L-1)A_{2}\text{,}  \nonumber
\\
E_{\pm } &=&\frac{\gamma _{L}+\gamma _{L-1}}{2}\pm \sqrt{\frac{(\gamma
_{L}-\gamma _{L-1})^{2}}{4}+\alpha _{L-1}\beta _{L}}\text{ }=\frac{\gamma
_{L}+\gamma _{L-1}}{2}\pm \sqrt{\frac{(\gamma _{L}-\gamma _{L-1})^{2}}{4}%
+8L(2L-1)A_{2}^{2}\text{ }}\text{.}  \nonumber
\end{eqnarray}
The similar procedure can be repeated for odd states.

\section{Generalization}

The obvious generalization of QES Bose Hamilatonians arises if Hamiltonian
itself does not have a ''canonical'' structure under description but can be
reduced to it with the help of the transformation $HK=KH^{\prime }$, where
the operator $K$ is some function of $a$ and $a^{+}$. In particular, it can
realize $u-v$ Bogolubov transformations. In what follows we assume that such
transformations, if needed, are already performed, so we try to generalize
such ''canonical'' forms themselves. Consider the action of Hamiltonian $%
H=H_{0}+V$ with $H_{0}$ from (\ref{h}) and

\begin{equation}
V=\sum_{s=0}^{s_{0}}%
\sum_{k=1}^{k_{0}}A_{sk}[(a^{+}a)^{s}a^{kq}+(a^{+})^{kq}(a^{+}a)^{s}]
\label{gen}
\end{equation}
on the functions $\Phi _{n}=x^{qn}$, where $q>0$ is an integer, $n=0$, $1$,
...:

\begin{eqnarray}
H_{x}\Phi _{n} &=&\gamma _{n}\Phi _{n}+\sum_{k=1}^{k_{0}}\alpha _{nk}\Phi
_{n+k}+\sum_{k=1}^{k_{0}}\beta _{nk}\Phi _{n-k}\text{,}  \label{genh} \\
\gamma _{n} &=&\sum_{p=1}^{p_{0}}\varepsilon _{p}(nq)^{p}\text{,}  \nonumber
\\
\alpha _{nk} &=&\sum_{s=0}^{s_{0}}A_{sk}(nq)^{s}\text{,}  \nonumber \\
\beta _{nk} &=&\sum_{s=0}^{s_{0}\text{ }%
}A_{sk}nq(nq-1)...(nq-kq+1)(nq-kq)^{s}\text{, }n\geq k\text{,}  \nonumber \\
\beta _{nk} &=&0\text{, }n<k\text{.}  \nonumber
\end{eqnarray}
Provided the conditions of cut off at $n=N$ are fulfilled, this Hamiltonian
can possess the invariant subspace $F=\{\Phi _{n}$, $n=0$, $1$, ...$N\}$, so
the wave function of the system $\Phi =\sum_{n=0}^{N}a_{n}\Phi _{n}$
includes the states with $nq$ ( $n=0$, $1$, ...$N$) particles only. The
conditions under discussion read now as follows. For any given $k$ we must
demand $\alpha _{N+1-i\text{, }k}=0$, $i=1$, ...$k$, so we have $k$
conditions. As $k=1$,$2$,...$k_{0}$, the total number of conditions is equal
to $n_{1}=\frac{(k_{0}+1)k_{0}}{2}$. On the other hand, the number of
coefficients $A_{sk}$ is equal to $n_{2}=(s_{0}+1)k_{0}$. The system can be
quasi-exactly solvable if $n_{2}>n_{1}$, so $2s_{0}\geq k_{0}$. In
particular, in accordance with examples considered above, one can always
adjust the coefficients properly, if $k_{0}=1$.

The approach considered in the present paper allows extension to
many-particle systems. In particular, for two pairs of Bose operators $%
(a,a^{+})$, $(b,b^{+})$ one may take 
\begin{equation}
H=\sum_{i}H_{i}^{(a)}h_{i}^{(b)}\text{,}  \label{b}
\end{equation}
where Hamiltonians $H_{i}$ and $h_{i}$ are built from operators $a$,$a^{+}$
and $b$, $b^{+}$, correspondingly, and have the structure (\ref{h}) or (\ref
{gen}).

\section{concluding remarks}

During recent years, the class of QES was extended considerably to include
two- and many-dimensional systems, matrix models \cite{ren}, the QES
anharmonic oscillator with complex potentials (\cite{uz}, p. 192; \cite
{compl}), etc. Meanwhile, it turned out that, apart from these (sometimes
rather sophisticated and exotic) situations, quasi-exact solvability exists
in an everyday life around us where anharmonic Bose oscillators can be met
at every step. In particular, the results obtained can be exploited in solid
state or molecular physics, theory of magnetism, etc. The approach suggested
in the present paper, shows the line along which a lot of second-quantized
models with algebraization of the part of the spectrum can be constructed.
This approach can be also extended to systems with interaction of subsystems
of different nature - in particular, between spin and Bose operator, Bose
and Fermi oscillators.





%
%

%
%

\end{document}